\documentstyle[12pt,epsf]{article}
\title{The Aharonov-Bohm scattering :\\
the role of the incident wave}
\author{
S. Sakoda\thanks{E-mail: sakoda@hc.keio.ac.jp}
\and
M. Omote\thanks{E-mail: omote@hc.keio.ac.jp} 
}
\def\boxit#1{ {\vbox{\hrule\hbox{\vrule\kern3pt
\vbox{\kern3pt{\vspace{1.8ex}#1}\kern3pt}\kern3pt\vrule}\hrule}}}
\makeatletter
\def\section{\@startsection{section}{1}{\z@}{3.5ex plus -1ex minus
 -.2ex}{2.3ex plus .2ex}{\large\bf}}
\def\subsection{\@startsection{subsection}{2}{\z@}
{3.25ex plus -1ex minus  -.2ex}
{1.5ex plus .2ex}{\normalsize\bf}}
\makeatother
\makeatletter
\@addtoreset{equation}{section}

\makeatother
\def\hcentering{%
	\oddsidemargin=211mm
	\advance\oddsidemargin-\textwidth
	\oddsidemargin=0.5\oddsidemargin
	\advance\oddsidemargin-1in
	\evensidemargin=\oddsidemargin}
\def\vcentering{%
	\topmargin=297mm
	\advance\topmargin-\textheight
	\topmargin=0.5\topmargin
	\headheight=0.4\topmargin
	\topmargin=0.2\topmargin
	\advance\topmargin-1in
	\headsep=\headheight
	\footskip=\headsep
	\footheight=\headheight}
\def\acknowledgments{\section*{ACKNOWLEDGMENTS}}

\newcommand{\nc}{\newcommand}

\def\itembox#1{\hbox to 3em{\hfill(#1)\hfill}}
\def\itembra#1{\hbox to 3em{\hfill[#1]\hfill}}
\def\eqref#1{(\ref{#1})}
\nc{\vphi}{\varphi}
\nc{\eps}{\epsilon}
\nc{\veps}{\varepsilon}

\makeatletter
\def\mathhexbox{\protect\mathhexbox@}
\def\mathhexbox@#1#2#3{\relax
\ifmmode\mathpalette{}{\m@th\mathchar"#1#2#3}%
\else\leavevmode\hbox{$\m@th\mathchar"#1#2#3$}\fi}
\def\text#1{%
\relax
\ifmmode %
\mathchoice {\hbox{\everymath{\displaystyle}\rm #1}}%
{\hbox{\everymath{\textstyle}\rm #1}}%
{\hbox{\everymath{\scriptstyle}%
\def\prm{\fam\z@ \the\scriptfont\z@ \relax}%
\def\pit{\fam\itfam \the\scriptfont\itfam \relax}%
\rm #1}%
}%
{\hbox{\everymath{\scriptscriptstyle}%
\def\prm{\fam\z@ \the\scriptscriptfont\z@ \relax}%
\def\pit{\fam\itfam \the\scriptscriptfont\itfam \relax}%
\rm #1}%
}%
\else %
\leavevmode\hbox{#1}%
\fi }
\def\bbox#1{%
\leavevmode\text{%
\textfont0 \the\textfont\bffam
\scriptfont0 \the\scriptfont\bffam
\scriptscriptfont0 \the\scriptscriptfont\bffam
\@temptokena\everymath \boldmath \everymath\@temptokena {$\m@th\relax#1$}%
}%
}
\font\fivbf=cmbx5 \font\sixbf=cmbx6 \font\sevbf=cmbx7 
\font\egtbf=cmbx8
\expandafter\def\expandafter\ixpt\expandafter{\ixpt
\scriptfont\bffam\sixbf \scriptscriptfont\bffam\fivbf}
\expandafter\def\expandafter\xpt\expandafter{\xpt
\scriptfont\bffam\sevbf \scriptscriptfont\bffam\fivbf}
\expandafter\def\expandafter\xipt\expandafter{\xipt
\scriptfont\bffam\egtbf \scriptscriptfont\bffam\sixbf}
\expandafter\def\expandafter\xiipt\expandafter{\xiipt
\scriptfont\bffam\egtbf \scriptscriptfont\bffam\sixbf}
\expandafter\def\expandafter\xivpt\expandafter{\xivpt
\scriptfont\bffam\tenbf \scriptscriptfont\bffam\sevbf}
\makeatother

\nc{\trac}[2]{{\textstyle\frac{#1}{#2}}}
\def\sgn#1{\mbox{\rm sgn}(#1)}
\def\dfrac#1#2{{{\displaystyle #1}\over{\displaystyle #2}}}

\def\derr{\partial_{r}}
\def\derrr{\partial_{r}^{2}}
\def\derv{\partial_{\vphi}}

\date{April 10, 1996}
\hcentering\vcentering
\begin{document}
\maketitle
\begin{center} {\em Department of Physics, Hiyoshi, Keio University, 
Hiyoshi, 
Yokohama 223, Japan}
\end{center}
\vspace{24pt}
\begin{abstract} 
The scattering problem under the influence of the Aharonov-Bohm (AB) 
potential is reconsidered. By solving the Lippmann-Schwinger (LS) 
equation we obtain the wave function of the scattering state in this 
system. In spite of working with a plane wave as an incident wave
we obtain the same wave function as was given by Aharonov and Bohm.
Another method to solve the scattering problem is given by making use 
of a modified version of Gordon's idea which was invented to consider
the scattering by the Coulomb potential.
These two methods give the same result, which guarantees the validity of 
taking an incident plane wave as usual to make an analysis of this scattering problem.
The scattering problem by a solenoid of finite radius is also 
discussed, and we find that the vector potential of the solenoid affects 
the charged particles even when the magnitude of the flux
is an odd integer as well as noninteger. It is shown that the unitarity of the $S$ matrix holds provided that a plane wave is taken to be an incident one.
\end{abstract}
\setcounter{page}{0}
\thispagestyle{empty}
\vfill\eject

\section{Introduction} 
\label{intro}

Since Aharonov and Bohm (AB) have discussed a scattering problem of a 
charged particle by a solenoid in order to clarify the significance 
of the vector potential in the quantum theory \cite{BoAh}, many people 
have considered the same problem from various viewpoints 
\cite{Takaba}--\cite{Alva}. 
As is well known, there are two approaches to deal with a 
scattering problem in the 
quantum theory. The first approach is to find a stationary state 
describing the scattering process by solving a time-independent 
Schr\"odinger equation. 
The second one is to study the time development of a 
wave packet with respect to a time-dependent Schr\"odinger equation. 
Most people as well as Aharonov and Bohm have analyzed 
the scattering by means of the first approach, and some people have 
discussed the same problem with the second approach \cite{Stelit}, \cite{Alva}.
As we see in the following, however, 
in spite of these efforts it seems not to be clear 
what is an incident wave in this scattering process.
In this paper we will try to answer this question 
with the first method because there seems to be lacking a common 
interpretation of the stationary wave function in the literature.

A system of charged particles interacting with the solenoid is  
described by a  Hamiltonian
\begin{equation}
\hat{H}={1\over{2\mu}}\left\{{\hat{\bbox{p}} -{e\over
c}\bbox{A}(\hat{\bbox{x}})}\right\}^{2}\ ,
\label{vecham1}
\end{equation}
where the electromagnetic vector potential $\bbox{A}$ is given by 
\[
\bbox{A}(\bbox{x})={\Phi\over2\pi r}
\left(-\sin\vphi,\cos\vphi\right),\qquad
(x,y)=(r\cos\vphi,r\sin\vphi).
\]
In order to study the scattering of the charged particles we 
solve the time-independent Schr\"odinger equation
\begin{equation}
	\hat{H}\psi_{E}(r,\vphi)=E\psi_{E}(r,\vphi)
	\label{schr1}
\end{equation}
to find an eigenfunction which describes the scattering process 
of charged particles.

Since the Hamiltonian commutes with the angular momentum, it can be 
easily shown that a most general solution of  \eqref{schr1} is given by  
\begin{equation}
	\psi_{E}(r,\vphi)=\sum_{n=-\infty}^{+\infty}
	c_{n}e^{in\vphi}J_{|n+\alpha|}(kr),\
	E={(\hbar k)^{2}\over 2\mu},
	\label{schsol1}
\end{equation}
where $J_{\nu}(x)$ denotes the Bessel function of $\nu$-th order 
and we have put $\alpha=-e\Phi/2\pi\hbar c$.
In \eqref{schsol1}, $c_{n}$'s are arbitrary constants to be determined by phisycal requirements.
Our main interest here and in the following is to ask 
what should be the correct choice for the coefficients 
$c_{n}$'s to describe the scattering process. 

It has been asserted by Aharonov and Bohm and many other 
authors that the incident wave for this scattering problem, when the 
incident beam comes into from the positive $x$ axis, should be a 
modulated plane wave $e^{-ikr\cos\varphi-i\alpha\varphi}$ to make the 
probability current of the incident wave constant. To fulfill 
this requirement the coefficients have been taken to be 
$c_{n}=(-i)^{|n+\alpha|}$. In the case of nonintegral $\alpha$, 
however, the incident wave becomes a multi-valued function. We 
may say that from the physical point of view 
it seems quite unsatisfactory to take such a multi-valued 
wave function as an incident one. 
We should also note that, for sufficiently large $r$, 
the term due to the vector potential 
do not contribute to the dominant part of
the current, even if we take a plane wave as an incident wave 
function.  
Thus, it is not conclusive to argue that the incident modulated wave 
gives the condition to determine those constants as $c_{n}=(-i)^{|n+\alpha|}$.

It will be, therefore, instructive 
to reconsider the scattering problem from other viewpoints. 
In this paper we try to 
find the wave function to describe the scattering process using two 
different ways.
The first one is to solve the Lippmann-Schwinger (LS) equation, 
which will be a standard method to consider 
the scattering problem of a quantum system,
taking a plane wave as an incident state instead of the modulated one. 
If we adopt the Born expansion to solve the LS equation, we will 
soon meet a difficulty because the perturbative method does not work 
for the present problem as is shown in \cite{AALL}, \cite{Nagel}.
Then we find an exact solution of the LS equation with the aid 
of the Feynman kernel.
The second method is an application of Gordon's idea 
which is proposed to discuss the scattering problem 
by the Coulomb potential \cite{Gordon}.
The idea may have been introduced to avoid the difficulty caused by 
the long-ranged nature of the 
Coulomb potential in formulating a scattering theory. 
This method will also be useful to examine scattering problems 
by other potentials with long-range effect.
It will be shown that these two methods give the
same wave function to describe the scattering state given by AB \cite{BoAh}.

If $\alpha=\text{integer}$, a further discussion is needed since the solution of the Schr\"odinger equation for infinitely thin 
solenoid does neither vanish 
nor be defined at the origin in that case. 
To ensure the 
impenetrability of the solenoid even for integral $\alpha$, we assume a solenoid to have a finite radius and consider this issue by generalizing the second method.

The plan of this paper is as follows. In section \ref{kernelkeisan} 
we give the Feynman kernel with effects of the solenoid 
potential using the path integral method. 
Section \ref{LSanalysis} is devoted to solving the LS equation exactly 
with the aid of the Feynman kernel, 
and the explicit form of the wave function will be obtained. 
In section \ref{gordon1} Gordon's method will be argued and its 
generalization to a system of a solenoid with finite radius will be 
done in section \ref{gordon2}. Conclusions and discussions in 
comparison with other's results are made in section \ref{discuss}.

\section{The Feynman kernel with effects of the solenoid potential} 
\label{kernelkeisan}
Since the complete set of the eigenfunctions 
\[
{1\over\sqrt{2\pi}}e^{in\varphi}J_{|n+\alpha|}(kr),\qquad n=0,\pm1,\pm2,\ldots
\]
for the Hamiltonian with effects of the solenoid are known,
we can immediately find an expression of the Feynman kernel 
\begin{equation} 
K(\bbox{x}_{F},\bbox{x}_{I};T)=\langle\bbox{x}_{F}\vert
e^{-i\hat{H}T/\hbar}
\vert\bbox{x}_{I}\rangle
\label{veckernel1}
\end{equation}
as its spectral representation \cite{OlaPop}, \cite{Stelit}
\begin{equation}
	K(\bbox{x}_{F},\bbox{x}_{I};T)=
	\int_{0}^{\infty}kdk e^{-i\hbar k^{2}T/2\mu}{1\over 2\pi}
	\sum_{n=-\infty}^{+\infty}e^{in(\vphi_{F}-\vphi_{I})}
	J_{|n+\alpha|}(kr_{F})J_{|n+\alpha|}(kr_{I}).
	\label{kernel0}
\end{equation}
To verify this expression, it suffices to notice the following facts, 
(i) it obeys the time-dependent Schr\"odinger equation, 
(ii) it is apparently single-valued with respect to both 
$\bbox{x}_{F}$ and $\bbox{x}_{I}$, 
(iii) it has the correct limit
\begin{equation}
\lim\limits_{t\rightarrow 0}K(\bbox{x}_{F},\bbox{x}_{I};t)=
\delta^{2}(\bbox{x}_{F}-\bbox{x}_{I}), 
\end{equation}
which follows from 
\begin{equation}
	{1\over \sqrt{ab}}\delta(a-b)=
	\int_{0}^{\infty}kdk J_{\nu}(ak)J_{\nu}(bk)\qquad
	[\text{Re}(\nu)>-1,\ a,b>0]\ ,
	\label{fourier-bessel}
\end{equation}
and
\begin{equation}
	\sum_{n=-\infty}^{+\infty}e^{in\theta}
	=2\pi\delta(\theta)\qquad [-\pi<\theta<\pi]\ .
	\label{deltatheta}
\end{equation}
If we carry out the integration with respect to $k$ in 
\eqref{kernel0}, we obtain 
\begin{equation}
	K(\bbox{x}_{F},\bbox{x}_{I};T)=
	{\mu \over2\pi i\hbar T}
	\sum_{n=-\infty}^{+\infty}\exp\left\{
	{i\mu \over2\hbar T}(r^{2}+r^{\prime}{}^{2})\right\} 
	I_{|n+\alpha|}\left({\mu rr^{\prime}\over i\hbar T}\right) 
	e^{in(\varphi-\varphi^{\prime})}\ .
	\label{kernel1}
\end{equation}

By use of \eqref{kernel0} or \eqref{kernel1} 
we can proceed to solve the LS equation.
In this section, however, we would like to give another derivation of 
\eqref{kernel0} because there seems to be some confusion in the path integral 
construction of the Feynman kernel in the 
literature \cite{BerIno}--\cite{Shiekh}.
Among them the most typical one would be the interpretation 
of its expression as {\em the sum over winding number\/} such as
\begin{equation}
	K(\bbox{x}_{F},\bbox{x}_{I};T)=\sum_{m=-\infty}^{+\infty}	
	e^{-i\alpha(\vphi_{F}-\vphi_{I}-2m\pi)}
	K_{m}(\bbox{x}_{F},\bbox{x}_{I};T),
	\label{winding}
\end{equation}
where $K_{m}(\bbox{x}_{F},\bbox{x}_{I};T)$ comes from paths going 
around the solenoid $m$ times in anticlockwise way, and the factor 
$e^{2\pi im\alpha}$ is a one dimensional representation of the fundamental 
group of the configuration space. 
In the following, however, we show that    
{\em the sum over winding number is not essential\/}, and that 
the sum in \eqref{winding} is to be interpreted as a result of the 
reformulation with aid of the Poisson sum formula 
for the expression obtained using the usual path integral method. 
To achieve this we will make use of the completeness relations of both 
eigenvectors $\vert\bbox{x}\rangle$ of $\hat{\bbox{x}}$ with 
eigenvalue $\bbox{x}$, and eigenvectors $\vert\bbox{p}\rangle$ of 
$\hat{\bbox{p}}$ with eigenvalue $\bbox{p}$, in formulating the 
path integral. As is easily recognized, the single-valuedness of 
\eqref{winding} is the consequence of these relations \cite{Ohnuki}.
 
Now we give the path integral derivation for the Feynman kernel of 
this system. Defining the exponential operator by
\begin{equation}
	e^{-i\hat{H}T/\hbar}=\lim\limits_{N\rightarrow\infty}
	\left(1-{i\veps\over\hbar}\hat{H}\right)^{N},\qquad
	\veps=T/N,
	\label{evopdef}
\end{equation}
and using the completeness of the states $\vert\bbox{x}\rangle$, we obtain
\begin{equation}
	K(\bbox{x}_{F},\bbox{x}_{I};T)=\lim\limits_{N\rightarrow\infty}
	\int\prod_{i=1}^{N-1}{d^{2}x(i)}\prod_{j=1}^{N}
	\langle\bbox{x}(j)\vert
\left(1-{i\veps\over\hbar}\hat{H}\right)
\vert\bbox{x}(j-1)\rangle .
	\label{kerneldef}
\end{equation}
Then, with the aid of the completeness of the states
$\vert\bbox{p}\rangle$,
we can express the infinitesimal version of the Feynman kernel as
\begin{eqnarray} 
& &
\langle\bbox{x}\vert
\left(1-{i\veps\over\hbar}\hat{H}\right)
\vert\bbox{x}^{\prime}\rangle
\nonumber\\ 
&=&
\lim\limits_{\delta\rightarrow0}\int{d^{2}p\over(2\pi\hbar)^{2}}
\exp\left\{{i\over\hbar}\bbox{p}(\bbox{x}-\bbox{x}^{\prime})
-{\delta\over2}\bbox{p}^2
\right\}
\left[1-{i\veps\over\hbar} {1\over{2\mu}}\left\{{\bbox{p} 
-{e\over c}\overline{\bbox{A}}
(\bbox{x},\bbox{x}^{\prime})}\right\}^{2}\right]\ 
,\label{hammat1}
\end{eqnarray}
where 
$\overline{\bbox{A}}(\bbox{x},\bbox{x}^{\prime})=
\{\bbox{A}(\bbox{x})+\bbox{A}(\bbox{x}^{\prime})\}/2$.

After shifting the integration variable $\bbox{p}$ by
\[
\bbox{p}\mapsto\bbox{p}+{e\over c}
\overline{\bbox{A}}(\bbox{x},\bbox{x}^{\prime}),
\]
we can rewrite \eqref{hammat1} as
\begin{eqnarray}
\langle\bbox{x}\vert
\left(1-{i\veps\over\hbar}\hat{H}\right)
\vert\bbox{x}^{\prime}\rangle
&=&
\exp\left\{{
{ie\over\hbar c}\overline{\bbox{A}}(\bbox{x},\bbox{x}^{\prime})
(\bbox{x}-\bbox{x}^{\prime})}\right\}
\nonumber\\ 
& &
\times
\lim\limits_{\delta\rightarrow0}\int{d^{2}p\over(2\pi\hbar)^{2}}
\exp\left\{{i\over\hbar}\bbox{p}(\bbox{x}-\bbox{x}^{\prime})
-{1\over2}\left(\delta+{i\veps\over\hbar\mu}\right)\bbox{p}^2
\right\}\ 
.\label{hammat2}
\end{eqnarray}
Carrying out the Gaussian integration with respect to  
$\bbox{p}$ and noting that
\begin{equation}
	\overline{\bbox{A}}(\bbox{x},\bbox{x}^{\prime})
(\bbox{x}-\bbox{x}^{\prime})
={\Phi\over4\pi}\left({r^{\prime}\over r}+{r\over r^{\prime}}\right)
\sin(\varphi-\varphi^{\prime}),
	\nonumber
\end{equation}
we obtain
\begin{eqnarray}
\langle\bbox{x}\vert\left(1-{i\varepsilon\over\hbar}\hat{H}\right)
\vert\bbox{x}^{\prime}\rangle & = &\lim\limits_{\delta\to 0}{\mu
e^{i\delta}\over2\pi i\hbar\varepsilon}
\exp\left[{i\mu e^{i\delta}\over2\hbar\veps}\{r^{2}+r^{\prime}{}^{2}
-2rr^{\prime}\cos(\varphi-\varphi^{\prime})\}
\right.
\nonumber\\ & &
\left. -{i\alpha\over2}\left({r^{\prime}\over r}+{r\over 
r^{\prime}}\right)
\sin(\varphi-\varphi^{\prime})\right],
\label{shortkernel1}
\end{eqnarray} 
where we have made a change of variables from Cartesian coordinate 
to the polar one and $\delta$ in \eqref{shortkernel1} has been 
renamed by
$1-i\hbar\mu\delta/\veps\mapsto e^{-i\delta}$.

To find the kernel for a finite time interval $T$ we need to perform $N-1$ integrations 
with respect to $\bbox{x}$'s in \eqref{kerneldef}. 
For this purpose, the form of the
exponent in \eqref{shortkernel1} is  extremely inconvenient. 
To overcome the difficulty it is useful to rewrite
\begin{eqnarray} 
&& 
{\mu 
rr^{\prime}\over\hbar\veps}e^{i\delta}\cos(\varphi-\varphi^{\prime})
+{\alpha\over2}\left({r^{\prime}\over r}+{r\over r^{\prime}}\right)
\sin(\varphi-\varphi^{\prime})
\nonumber\\ 
& = &
\left\{\left({\mu rr^{\prime}\over\hbar\veps}e^{i\delta}\right)^{2}
+{\alpha^{2}\over4}\left({r^{\prime}\over r}+{r\over  
r^{\prime}}\right)^{2}
\right\}^{1/2}\cos(\varphi-\varphi^{\prime}-\theta_{\veps})
\nonumber\\ 
& = & 
{\mu 
rr^{\prime}\over\hbar\veps}e^{i\delta}\sqrt{1+\tan^{2}\theta_{\veps}}
\cos(\varphi-\varphi^{\prime}-\theta_{\veps}),
\end{eqnarray} 
where
\begin{equation}
\tan\theta_{\veps}={\hbar\veps\alpha\over2\mu rr^{\prime}}e^{-i\delta}
\left({r^{\prime}\over r}+{r\over r^{\prime}}\right).
\end{equation}
Upon integration with respect to $\bbox{x}$ or $\bbox{x}^{\prime}$, 
the Gaussian part in the integrand will dominate for sufficiently 
small $\veps$. We may, therefore, regard components of 
$\bbox{x}-\bbox{x}^{\prime}$ as $O(\veps^{1/2})$. Then it follows that
\begin{equation}
{1\over2}\left({r^{\prime}\over r}+{r\over r^{\prime}}\right)
=1+O(\veps^{1/2}).
\end{equation} 
(Note, however, 
that the same argument does not hold true for $\varphi-\varphi^{\prime}$.)
Recalling that we may discard terms of $O(\veps^{\rho})$ for $\rho>1$, 
in the exponent of a path integral, we can replace the definition of 
$\tan\theta_{\veps}$ by
\[
\tan\theta_{\veps}=
{\hbar\veps\alpha\over\mu rr^{\prime}}e^{-i\delta}
\left\{1+O(\veps^{1/2})\right\}
=\theta_{\veps}\left\{1+O(\veps^{1/2})\right\}.
\]

Thus we obtain
\begin{eqnarray} 
&&
\exp\left[ -i{\mu rr^{\prime}\over\hbar\veps}
e^{i\delta}\cos(\varphi-\varphi^{\prime})
-{i\alpha\over2}\left({r^{\prime}\over r}+{r\over r^{\prime}}\right)
\sin(\varphi-\varphi^{\prime})
\right]
\nonumber\\
& = &
\exp\left[ -i{\mu rr^{\prime}\over\hbar\veps}
e^{i\delta}\sqrt{1+\theta_{\veps}^{2}}
\cos(\varphi-\varphi^{\prime}-\theta_{\veps})\right]
\left\{1+O(\veps^{3/2})\right\}
\nonumber\\ 
& = &
\sum_{n=-\infty}^{+\infty} I_{|n|}\left({\mu rr^{\prime}\over 
i\hbar\veps}e^{i\delta}\sqrt{1+\theta_{\veps}^{2}}\right)
e^{in(\varphi-\varphi^{\prime}-\theta_{\veps})}
\left\{1+O(\veps^{3/2})\right\}
\ .
\label{pathact}
\end{eqnarray} 
When $\veps$ becomes small, the argument of the modified Bessel 
function grows to allow us to apply its asymptotic form. Then 
keeping terms up to $O(\veps)$ in the exponent, we obtain
\begin{eqnarray} 
&&
I_{|n|}\left({\mu rr^{\prime}\over 
i\hbar\veps}e^{i\delta}\sqrt{1+\theta_{\veps}^{2}}\right)
e^{-in\theta_{\veps}}
\nonumber\\ 
&=&
\sqrt{{i\hbar\veps\over2\pi\mu
rr^{\prime}}e^{-i\delta}}
\exp\left[ {\mu rr^{\prime}\over i\hbar\veps}
e^{i\delta}-{i\hbar\veps\over2\mu 
rr^{\prime}}e^{-i\delta}
\{(n+\alpha)^{2}-1/4\}
\right]\left\{1+O(\veps^{3/2})\right\}
\nonumber\\ 
&=&
I_{|n+\alpha|}\left({\mu rr^{\prime}\over 
i\hbar\veps}e^{i\delta}\right)
\left\{1+O(\veps^{3/2})\right\}.
\label{besselch}
\end{eqnarray} 
Substituting it into \eqref{pathact}, we arrive at
\begin{eqnarray} 
&&
\exp\left[ -i{\mu rr^{\prime}\over\hbar\veps}
e^{i\delta}\cos(\varphi-\varphi^{\prime})
-{i\alpha\over2}\left({r^{\prime}\over r}+{r\over r^{\prime}}\right)
\sin(\varphi-\varphi^{\prime})
\right]
\nonumber\\
& = &
\sum_{n=-\infty}^{+\infty} I_{|n+\alpha|}\left({\mu rr^{\prime}\over 
i\hbar\veps}e^{i\delta}\right)
e^{in(\varphi-\varphi^{\prime})}
\left\{1+O(\veps^{3/2})\right\}
\ .
\label{pathact1}
\end{eqnarray} 
Therefore the infinitesimal kernel
\eqref{shortkernel1} is now  rewritten as
\begin{eqnarray} 
K(\bbox{x},\bbox{x}^{\prime};\veps) 
&=&
\langle\bbox{x}\vert\left(1-{i\veps\over\hbar}\hat{H}\right)
\vert\bbox{x}^{\prime}\rangle
\nonumber\\ 
& = &
\lim\limits_{\delta\rightarrow0}{\mu e^{i\delta}\over2\pi i\hbar\veps}
\sum_{n=-\infty}^{+\infty}\exp\left\{ -{\mu
e^{i\delta}\over2i\hbar\veps}(r^{2}+r^{\prime}{}^{2})
+in(\varphi-\varphi^{\prime})\right\}
\nonumber\\ 
&&
\mathop{\hphantom{
\lim\limits_{\delta\rightarrow0}
{\mu e^{i\delta}\over2\pi i\hbar\veps}
\sum_{n=-\infty}^{+\infty} }}\times 
I_{|n+\alpha|}\left({\mu rr^{\prime}\over
i\hbar\veps}e^{i\delta}\right)
\left\{1+O(\veps^{3/2})\right\}.
\label{shortkernel4}
\end{eqnarray} 
Here a comment is in need; in obtaining the result of 
\eqref{besselch} we have discarded the possibility to use the 
modified Bessel function of negative order since it breaks the 
regularity of the kernel at the origin.
 
It is now straightforward to see that the multiplication rule holds:
\begin{equation}
\int d^{2}\bbox{x} K(\bbox{x}_{2},\bbox{x};\veps)
K(\bbox{x},\bbox{x}_{1};\veps) =
K(\bbox{x}_{2},\bbox{x}_{1};2\veps),
\label{multiplication}
\end{equation} 
since the integration with respect to the angle variable is
trivial and we may  make use of a formula
\begin{equation}
\int_{0}^{\infty}{rdr}e^{-ar^{2}}J_{\mu}(pr)J_{\mu}(qr) ={1\over
2a}e^{-(p^{2}+q^{2})/4a}I_{\mu}(pq/2a),
\end{equation} 
which holds for $|{\rm arg}(a)|<\pi/2,\ {\rm Re}(\mu)>-1,\ p,q>0$.
Repeated use of the rule
\eqref{multiplication}  (and putting all $\delta$'s to $0$ after
integration) will lead us to 
\begin{equation} 
K(\bbox{x},\bbox{x}^{\prime};T) ={\mu \over2\pi i\hbar T}
\sum_{n=-\infty}^{+\infty}\exp\left\{ {i\mu \over2\hbar
T}(r^{2}+r^{\prime}{}^{2})\right\} 
I_{|n+\alpha|}\left({\mu rr^{\prime}\over
i\hbar T}\right) e^{in(\varphi-\varphi^{\prime})}.
\label{fkernel1}
\end{equation} 
Thus \eqref{kernel1} is again obtained by the usual formulation of 
path integral. Here it should be noticed that the sum over winding 
numbers in formulating the path integral is not essential.

\section{The wave function of a scattering state 
as a solution of Lippmann-Schwinger equation} 
\label{LSanalysis}
In this section we obtain the wave function for the scattering state 
of charged particles scattered by the solenoid.
It is known that, for the present problem, 
the Born approximation fails to give a reliable answer 
because we cannot avoid a divergent integral
$
\int_{0}^{r} dr'J_{0}^{2}(kr')/r'
$
even in its first order \cite{AALL}, \cite{Nagel}. The iterative 
method to solve the LS equation will also be unsatisfactory by the 
same reason.
Therefore we need to solve it in an exact way
with the aid of the Feynman kernel given in \eqref{fkernel1}. 
 
The LS equation for the system reads
\begin{eqnarray}
\psi_E\left( {r,\varphi } \right) 
& = & u_E\left( {r,\varphi }\right)+
\psi_{\text{S}}\left( {r,\varphi }\right),
\label{LSeq}\\
\psi_{\text{S}}\left( {r,\varphi }\right) 
&=&
\int_0^\infty  {r'dr'}
\int_{-\pi}^{+\pi }  {d\varphi  '}\left\langle \bbox{x}
\right|\left( {E-\hat{H}+i\varepsilon } 
\right)^{-1}\left| 
\bbox{x'} \right\rangle  {{\hbar ^2} \over {2\mu r'^2}}
\alpha\left( {-2i\partial_{\varphi '}+\alpha } \right)
u_E\left( {r',\varphi '} \right)\ ,
\nonumber
\end{eqnarray} 
where we have taken an incident plane wave 
$u_E\left( {r,\varphi } \right)=e^{ikr\cos\theta}\ 
(\theta=\varphi -\varphi_0)$ as an eigenstate of the free 
Hamiltonian and $\varphi_0$ indicates the 
direction of the incident beam. From \eqref{fkernel1}, we can easily 
obtain the Green's function in the above by Laplace transform
\begin{equation}
\left\langle \bbox{x} \right|\left( {E-\hat{H}+i\varepsilon } 
\right)^{-1}\left| \bbox{x'} \right\rangle \nonumber\\ =
\int_0^\infty  {{{dT} 
\over {i\hbar }}e^{{{i\left( {E+i\varepsilon } \right)T} 
\mathord{\left/ {\vphantom {{i\left( {E+i\varepsilon } \right)T} 
\hbar }} \right. \kern-\nulldelimiterspace} \hbar }}\left\langle 
\bbox{x} 
\right|e^{{{-i\hat{H}T} \mathord{\left/ {\vphantom {{-i\hat{H}T} 
\hbar }}
\right. 
\kern-\nulldelimiterspace} \hbar }}\left| \bbox{x'} \right\rangle  }.
\end{equation} 
By putting $E=\hbar^{2}k^{2}/2\mu$, it turns out to be
\begin{eqnarray}
\left\langle \bbox{x} \right|\left( {E-\hat{H}+i\varepsilon } 
\right)^{-1}\left| \bbox{x'} \right\rangle 
&=& {\mu  \over {2i\hbar ^2}}
\sum\limits_{n=-\infty }^{+\infty }  e^{in\left( {\varphi -\varphi '}
\right)}\left\{ {\theta \left(  {r-r'} \right)H_{\left| {n+\alpha }
\right|}^{\left( 1 \right)}\left(  {kr} \right)J_{\left| {n+\alpha }
\right|}\left( {kr'} \right) }\right.\nonumber\\ 
&&
\mathop{\hphantom{{\mu  \over {2i\hbar ^2}}
\sum\limits_{n=-\infty  }^{+\infty } 
}}
\left.{ +\theta\left( {r'-r} \right)J_{\left| {n+\alpha } 
\right|}\left( {kr} 
\right)H_{\left| {n+\alpha } \right|}^{\left( 1 \right)}\left( {kr'} 
\right)} \right\},
\label{LSkernel}
\end{eqnarray} 
where use has been made of a formula
\begin{equation}
	\int_{0}^{\infty}pdp{J_{\nu}(ap)J_{\nu}(bp)\over{p^{2}-k^{2}-i\eps}}
	={\pi i\over 2}H_{\nu}^{(1)}(ak)J_{\nu}(bk)\qquad
	[\text{Re}(\nu)>-1,\ a\ge b>0],
	\label{besselkousiki}
\end{equation}
and $\theta(x)$ is the step function.

Substituting \eqref{LSkernel} and partial wave expansion of the plane 
wave $u_E\left( {r,\varphi } \right)$ into the integrand of 
$\psi_{\text{S}}\left( {r,\varphi } \right)$, we obtain
\begin{equation}
\psi_{\text{S}}\left( {r,\varphi }\right) =
\sum\limits_{n=-\infty }^{+\infty } 
{\left\{ {A_n\left( r \right)
H_{\left| {n+\alpha } \right|}^{\left( 1 \right)}\left( {kr} \right)
+B_n\left( r \right)J_{\left| {n+\alpha } \right|}\left( {kr} 
\right)} \right\}
e^{in\theta +{{i\left|  n \right|\pi } 
\mathord{\left/ {\vphantom {{i\left| n \right|\pi }  2}} \right.
\kern-\nulldelimiterspace} 2}}}\ ,
\label{LSeq2}
\end{equation} 
where
\begin{eqnarray} 
A_n\left( r \right)
& = &
{\pi  \over {2i}}\alpha \left( 
{2n+\alpha } \right)\int_0^r {{{dr'} \over {r'}}J_{\left|  {n+\alpha }
\right|}\left( {kr'} \right)J_{\left| n \right|}\left(  {kr'} 
\right)}\ ,
\nonumber\\ 
B_n\left( r \right)
& = &
{\pi  \over {2i}}\alpha \left( {2n+\alpha } 
\right)\int_r^\infty  {{{dr'} \over {r'}}H_{\left| {n+\alpha } 
\right|}^{\left( 1 \right)}\left( {kr'} \right)J_{\left| n 
\right|}\left( {kr'} \right)}.
\label{keisuu}
\end{eqnarray} 
Making use of a formula of indefinite integral for cylindrical 
functions(represented by $Z_{\mu}$ and $\tilde Z_{\nu}$ for the sake of convenience)
\begin{eqnarray}
\int {{{dx} \over x}Z_\mu \left( {ax} \right)
\tilde Z_\nu \left(  {ax} \right)}
& = & 
-{{ax} \over {\mu ^2-\nu ^2}}
\left\{ {Z_{\mu +1}\left(  {ax} \right)\tilde
Z_\nu \left( {ax} \right)-Z_\mu \left( {ax} 
\right)\tilde Z_{\nu +1}\left( {ax} \right)} \right\}
\nonumber\\ 
& & 
+{{Z_\mu \left({ax} \right)\tilde Z_\nu \left( {ax} \right)} 
\over {\mu +\nu}},\quad \left( {\mu \ne \nu } \right),
\end{eqnarray} 
we obtain
\begin{eqnarray}
A_{n}(r)
& = &
{{i\pi} \over2}kr\left\{ 
{J_{\left| {n+\alpha } \right|+1}\left( {kr} \right)J_{\left| n 
\right|}\left( {kr} \right)-J_{\left| {n+\alpha } \right|}\left( {kr} 
\right)J_{\left| n \right|+1}\left( {kr} \right)} 
\right\}
\nonumber\\
& &
-{{i\pi} \over2}{{\alpha \left( {2n+\alpha }\right)} \over {\left| 
{n+\alpha } 
\right|+\left| n \right|}}J_{\left| {n+\alpha } \right|}
\left( {kr} \right)J_{\left| 
n \right|}\left( {kr} \right)
\label{keisuua}
\end{eqnarray}
and
\begin{eqnarray}
B_{n}(r)
& = &
-{{i\pi} \over2}kr
\left\{ {H_{\left| {n+\alpha } \right|+1}^{\left( 1 
\right)}\left( {kr} \right)J_{\left| n \right|}\left( {kr} 
\right)-H_{\left| {n+\alpha } \right|}^{\left( 1 \right)}\left( {kr} 
\right)J_{\left| n \right|+1}\left( {kr} \right)} 
\right\}\nonumber\\
& &
+{{i\pi} \over2}{{\alpha \left( {2n+\alpha }\right)} \over {\left| 
{n+\alpha } 
\right|+\left| n \right|}}
H_{\left| {n+\alpha } \right|}^{\left( 1 \right)}\left({kr} \right)
J_{\left| n \right|}\left( {kr} \right)
+e^{-i(|n+\alpha|- n )\pi/2 } .
\label{keisuub}
\end{eqnarray}
 
By a simple calculation with the aid of Lommel's formula
\[
J_{\nu+1}(x)H_{\nu}^{(1)}(x)-J_{\nu}(x)H_{\nu+1}^{(1)}(x)={2i\over\pi 
x}\ ,
\]
we have
\begin{eqnarray}
& &
\left\{ {A_n\left( r \right)
H_{\left| {n+\alpha } \right|}^{\left( 1 \right)}\left( {kr} \right)
+B_n\left( r \right)J_{\left| {n+\alpha } \right|}\left( {kr} 
\right)} \right\}
e^{in\theta+{{i\left| n \right|\pi } 
\mathord{\left/ {\vphantom {{i\left| n \right|\pi } 2}} 
\right. \kern-\nulldelimiterspace} 2}}\nonumber\\
& = &
-J_{\left| n \right|}\left( {kr} \right)
e^{in\theta+{{i\left| n \right|\pi } 
\mathord{\left/ {\vphantom {{i\left| n \right|\pi } 2}} \right. 
\kern-\nulldelimiterspace} 2}}
+J_{\left| {n+\alpha } \right|}\left( {kr} \right)
e^{in\theta+i n\pi -{{i\left| {n+\alpha } \right|\pi } 
\mathord{\left/ {\vphantom {{i\left| {n+\alpha } \right|\pi } 2}} 
\right. 
\kern-\nulldelimiterspace} 2}}\ .
\end{eqnarray}
Then \eqref{LSeq2} can be rewritten as
\begin{equation}
\psi_{\text{S}}\left( {r,\varphi }\right) 
=-e^{ikr\cos \theta}
+\sum\limits_{n=-\infty }^{+\infty } 
{J_{\left| {n+\alpha }\right|}\left( {kr} \right)
e^{in\theta+in\pi -{{i\left| {n+\alpha } \right|\pi } 
\mathord{\left/ {\vphantom {{i\left| {n+\alpha } \right|\pi } 2}} 
\right. \kern-\nulldelimiterspace} 2}}}.
	\label{psiscatt}
\end{equation} 
We thus find that the total wave function for the scattering state is 
given by
\begin{equation}
\psi _E\left( {r,\varphi } \right)
=\sum\limits_{n=-\infty  }^{+\infty }
{J_{\left| {n+\alpha } \right|}\left( {kr} \right)
e^{in\theta+in\pi  -{{i\left| {n+\alpha } \right|\pi } 
\mathord{\left/ {\vphantom {{i\left| {n+\alpha } \right|\pi } 2}}
\right. 
\kern-\nulldelimiterspace} 2}}}.
\label{wholepsi}
\end{equation} 
It is very interesting to recognize that by putting $\varphi_{0}=\pi$ 
the solution of LS equation coincides with 
the wave function obtained by AB \cite{BoAh}. 
But we have to remember that in solving the LS equation
we take the plane wave as an incident wave and that the resulting 
scattered wave is given by \eqref{psiscatt}. 
In spite of the fact that the total wave 
function is same as that of AB,
both the incident wave and the scattered wave in this section are 
different from those of AB as a consequence.

Next we proceed  to find the differential cross section by use of 
the scattered wave \eqref{psiscatt}. 
In view of \eqref{keisuua} and \eqref{keisuub}, we 
notice that
$A_{n}(r)=O((kr)^{0})$ while 
$B_{n}(r)=O((kr)^{-1})$ for large $kr$.
Then we easily obtain the asymptotic form of scattered
wave $\psi_{\text{S}}\left( {r,\varphi }\right)$ 
from \eqref{LSeq2} as
\begin{equation}
\psi_{\text{S}}\left( {r,\varphi } \right)\mathop \sim 
\limits_{r\to \infty}
\sum_{n=-\infty}^{+\infty}A_{n}(\infty)H_{|n+\alpha|}^{(1)}(kr)
e^{in\theta+i|n|\pi/2},
\label{LSscasymp}
\end{equation} 
where $A_{n}(\infty)$ is found from \eqref{keisuua} to be
\begin{equation} 
A_{n}(\infty)=-i\sin\{(|n+\alpha|-|n|)\pi/2\}.
\label{Aninfty}
\end{equation} 
Using the asymptotic form of the Hunkel functions 
and \eqref{Aninfty} for $A_{n}(\infty)$ in \eqref{LSscasymp},
we are lead to
\begin{equation}
	\psi_{\text{S}}\left( {r,\varphi } \right)\sim
	{1\over\sqrt{2\pi kr}}e^{ikr-i\pi/4}
	\sum_{n=-\infty}^{+\infty}
	(e^{2i\delta_{n}(\alpha)}-1)e^{in\theta},
	\label{LSscasymp2}
\end{equation}
where the phase-shift in $n$-th partial wave is given by
\begin{equation}
	\delta_{n}(\alpha)=
	\left\{{
	\begin{array}{cc}
		-\pi\alpha/2 & (n+[\alpha]\ge 0)  \\
		+\pi\alpha/2 & (n+[\alpha]<0)
	\end{array}}\right.\ .
	\label{phaseshift}
\end{equation}
Here and in the following we denote the integral part of $\alpha$ 
by $[\alpha]$ and its nonintegral part by $\{\alpha\}$ to write 
$\alpha=[\alpha]+\{\alpha\}$.

We here introduce a regularization parameter $\eps$ for the sum in 
\eqref{LSscasymp2} 
so that it is defined as an Abel sum because the phase-shift does not 
decrease 
at all when $|n|$ becomes large and define $f(\theta)$ as 
 \begin{equation}
 	f(\theta)=\lim\limits_{\eps\rightarrow 0}
 	{e^{-i[\alpha]\theta}\over\sqrt{2\pi k}}
 	\left\{{
 	\sum_{n=0}^{+\infty}(e^{-i\pi\alpha}-1)e^{in\theta-n\eps}
 	+\sum_{n=1}^{+\infty}(e^{i\pi\alpha}-1)e^{-in\theta-n\eps}
 	}\right\}\ .
 	\label{scamp1}
 \end{equation}
Then performing the sum of geometric series and making use 
of a symbolic relation
\[
\lim\limits_{\eps\rightarrow 0}{1\over {x-a\pm i\eps}}
=P{1\over{x-a}}\mp i\pi\delta(x-a)
\]  
with denoting the principal value by $P$, 
we finally obtain
in terms of the scattering amplitude $f(\theta)$
\begin{eqnarray}
\psi_{\text{S}}\left( {r,\varphi } \right)
& \sim & 
{1\over\sqrt{r}}e^{ikr-i\pi/4}f(\theta)
\nonumber\\
f(\theta)
& = &
\sqrt{{2\pi\over k}}
\left\{(\cos\pi\alpha-1)\delta(\theta)
+i{\sin\pi\alpha\over\pi}
e^{-i[\alpha]\theta}
P{1
\over e^{i\theta}-1}\right\}.
\label{psasymp1}
\end{eqnarray}

Although the total wave function has happened to have exactly same form as  
the result of AB as mentioned in the above, the scattering 
amplitude \eqref{psasymp1} disagrees to 
that of AB \cite{BoAh} or \cite{Takaba} by the 
$\delta$ function term. 
Nevertheless, this disagreement can be discarded when we have an interest in 
the differential cross section only for non-forward direction($\theta\ne 0$) 
since we cannot well separate the scattered and un-scattered particles 
in the forward direction experimentally. As far as in the  
non-forward direction, the differential cross section 
is thus given by
\begin{equation}
	d\sigma(\alpha)=
	{1\over 2\pi k}{\sin^{2}\pi\alpha\over\sin^{2}(\theta/2)}
	d\theta\ .
	\label{difcs}
\end{equation}
In this sense the scattering amplitude of AB describes the physics 
appropriately. However, 
if we take into account 
the unitarity of the $S$-matrix, 
the $\delta$ function for the forward direction cannot be 
neglected as is pointed out by Ruijsenaars \cite{Ruijs}. 
A more detailed description of the property of the $S$-matrix for this 
system is given in appendix \ref{smatrix}.

\section{Another derivation of the scattering state} 
\label{gordon1}

We here consider the other approach to the problem by using a 
modified version of 
Gordon's idea which has been proposed
in the analysis for the
scattering of a charged particle by the Coulomb potential \cite{Gordon}. 
The essence of the method is to prepare the asymptotic region described 
by the free Hamiltonian far
distant from the solenoid in order to overcome some difficulties 
caused by the long range effects of the solenoid field. 
To this aim we introduce a modified vector potential
\begin{equation}
\bbox{A}=
\left\{
\begin{array}{cc}
\dfrac{\Phi}{2\pi}\left(\dfrac{1}{r_{0}^2}-
\dfrac{1}{R^2}\right)r\bbox{e}_{\vphi}& (0\le r\le r_{0})\\
\dfrac{\Phi}{2\pi}\left(\dfrac{1}{r}-
\dfrac{r}{R^2}\right)\bbox{e}_{\vphi}& (r_{0}< r\le R)\\ 
0& (R<r)
\end{array}
\right. ,
\end{equation} 
where $r_{0}$ is the radius of the shielded solenoid. It should be 
noticed that in the region $R<r$ the vector potential does not affect
charged particles. To go back to the original AB problem
we just put $R\rightarrow\infty$ after solving 
the Schr\"odinger equation for this system. In this section, we first deal
with the scattering by an infinitely thin solenoid($r_{0}=0$), and 
then we generalize the analysis to the case of a finite 
size($r_{0}>0$) solenoid in the next section.

In the asymptotic region($r>R$) where the vector potential 
is absent, the solution of the Schr\"odinger equation is given 
by eigenstates of the free Hamiltonian 
and the wave function to describe the scattering state 
$\psi_{\text{II}}(r,\vphi)$ 
will be given by
\begin{equation}
\psi_{\text{II}}(r,\vphi)
=e^{ikr\cos\theta}+
\sum_{n=-\infty}^{+\infty}
a_{n}H_{n}^{(1)}(kr)e^{in\theta}\quad
(\theta=\vphi-\vphi_{0}),
\label{psi2}
\end{equation}
where $a_{n}$'s are constant coefficients to be determined in the following. 
In the scattering region $(0 <r \le R)$ the wave function 
$\psi_{\text{I}}(r,\vphi)$
is subjected to
\begin{eqnarray}
\hat{H}_{\text{I}}\psi_{\text{I}}(r,\vphi)
&=&{\hbar^2 k^2\over 2\mu}\psi_{\text{I}}(r,\vphi),\\
\hat{H}_{\text{I}}&=&
-{\hbar^2 \over 2\mu}\left[{
\derrr+{1\over r}\derr+{1\over r^2}
\left\{{
\derv+i\alpha\left({
1-{r^2\over R^2}
}\right)
}\right\}^2
}\right].
\end{eqnarray}
Assuming the partial wave expansion for $\psi_{\text{I}}(r,\vphi)$
\begin{equation}
\psi_{\text{I}}(r,\vphi)=
\sum_{n=-\infty}^{+\infty}e^{in\theta}\psi_{\text{I},n}(r),
\end{equation}
\begin{equation}
\left[{
\derrr+{1\over r}\derr-{1\over r^2}
\left\{{
n+\alpha\left({
1-{r^2\over R^2}
}\right)
}\right\}^2
+k^2
}\right]
\psi_{\text{I},n}(r)=0
\end{equation}
and making a change of variable $r\mapsto z=\alpha (r/R)^2$ with
$\psi_{\text{I},n}=W_{n}/\sqrt{z}$, we obtain
\begin{equation}
W^{\prime\prime}_{n}+\left\{{
-{1\over 4}+{\lambda+\nu/2\over z}
-{(\nu/2)^2-1/4\over z^2}
}\right\}W_{n}=0.
\label{whittakereq}
\end{equation}
In the above we have put $\nu=n+\alpha$, $\lambda=(kR)^2/(4\alpha)$,
and a prime denotes the differentiation with respect to $z$. 
The general solution of \eqref{whittakereq} is given by a linear 
combination of the Whittaker functions
\begin{equation}
W_{n}(z)=b_{n} M_{\lambda+\nu/2,\ |\nu|/2}(z)
+c_{n} M_{\lambda+\nu/2,\ -|\nu|/2}(z),
\label{whittakersol}
\end{equation}
where $b_{n}$ and $c_{n}$ are arbitrary constants and $M_{\kappa,\ 
\mu}(z)$ is defined by
\begin{eqnarray}
M_{\kappa,\ \mu}(z)
& = &
z^{\mu+1/2}e^{-z/2}{}_{1}F_{1}(\mu-\kappa+1/2;2\mu+1;z)
\nonumber\\
& = &
z^{\mu+1/2}e^{-z/2}
\sum_{l=0}^{\infty}{\Gamma(2\mu+1)\Gamma(\mu-\kappa+l+1/2)\over
\Gamma(2\mu+l+1)\Gamma(\mu-\kappa+1/2)}{z^{l}\over l!}.
\label{whitfunc}
\end{eqnarray}
The regularity of the wave function at the origin implies that the 
coefficient $c_{n}$ of the singular solution $M_{\lambda+\nu/2,\ -|\nu|/2}(z)$ 
in \eqref{whittakersol} must 
vanish. Therefore the solution for the scattering region is given by
\begin{equation}
\psi_{\text{I}}(r,\vphi)=
\sum_{n=-\infty}^{+\infty}e^{in\theta}{b_{n}\over\sqrt{z}}
M_{\lambda+\nu/2,\ |\nu|/2}(z).
\label{psi1}
\end{equation}

To determine the coefficients $a_{n}$ in \eqref{psi2} and $b_{n}$ in 
\eqref{psi1}, we require the continuity of the wave
function itself as well as its derivative in the normal direction on the 
surface $r=R$.
These conditions may be imposed on each partial wave independently 
to give
\begin{eqnarray}
	a_{n}(R) & = & {1\over2}e^{in\pi/2}
	\left\{{{\bar{p}_{n}(R)\over p_{n}(R)}-1}\right\}\ ,
	\label{keisuuan1}  \\
	b_{n}(R) & = & 
	{\sqrt{\alpha}\over2}e^{in\pi/2}{H_{n}^{(2)}(kR)
	+e^{2i\delta_{n}(R;\alpha)}H_{n}^{(1)}(kR)
	\over M_{\lambda+\nu/2,\ |\nu|/2}(\alpha)}\ .
	\label{keisuubn1}
\end{eqnarray}
Here $p_{n}(R)$ in \eqref{keisuuan1} is defined by
\begin{eqnarray}
	p_{n}(R) & = & kR\left\{H_{n-1}^{(1)}(kR)-H_{n+1}^{(1)}(kR)\right\}
	M_{\lambda+\nu/2,\ |\nu|/2}(\alpha)
	\nonumber\\
	& &
	-2H_{n}^{(1)}(kR)\left\{(\alpha-2\lambda-\nu-1)
	M_{\lambda+\nu/2,\ |\nu|/2}(\alpha)
	\right.
	\nonumber\\
	& &
	\left.
	+(|\nu|+2\lambda+\nu+1)M_{\lambda+\nu/2+1,\ |\nu|/2}(\alpha)
	\right\}
	\label{qkansu}
\end{eqnarray}
and we have put $e^{2i\delta_{n}(R;\alpha)}=\bar{p}_{n}(R)/ 
p_{n}(R)$ because it should be identified with the 
phase-shift of $n$-th partial wave. Thus, we obtain a solution in the 
scattering region 
\begin{equation}
	\psi_{\text{I}}(r,\vphi)=\sum_{n=-\infty}^{+\infty}
	{1\over2}e^{in\pi/2}{H_{n}^{(2)}(kR)
	+e^{2i\delta_{n}(R;\alpha)}H_{n}^{(1)}(kR)
	\over M_{\lambda+\nu/2,\ |\nu|/2}(\alpha)}
	{R\over r}
	M_{\lambda+\nu/2,\ |\nu|/2}(\alpha r^{2}/R^{2})
	\label{psiofRi}
\end{equation}	
and also in the asymptotic region
\begin{equation}	
	\psi_{\text{II}}(r,\vphi)=e^{ikr\cos\theta}
	+{1\over2}\sum_{n=-\infty}^{+\infty}
	H_{n}^{(1)}(kr)\left\{e^{2i\delta_{n}(R;\alpha)}-1\right\}\ .
	\label{psiofRo}
\end{equation}

We put $R\rightarrow\infty$ in 
\eqref{psiofRi} and \eqref{psiofRo} by making use of 
the well-known asymptotic forms of cylindrical functions 
and that of the Whittaker function
\begin{equation}
M_{\kappa,\ \mu}(z)\sim
{1\over\sqrt{\pi}}\Gamma(1+2\mu)\kappa^{-\mu-1/4}z^{1/4}
\cos(2\sqrt{\kappa z}-\mu\pi-\pi/4),
\end{equation}
for $\text{Re}(\kappa)\gg |z|,|\mu|,\ \text{Re}(z)>0$ 
in \eqref{qkansu} to obtain the phase-shift
\[
\delta_{n}(\infty;\alpha)=-(|\nu|-n)\pi/2\ .
\] 
Then the coefficients of scattered wave in the asymptotic region is 
found to be
\begin{equation}
a_{n}(\infty)={1\over2}e^{in\pi/2}\left\{e^{-i(|\nu|-n)\pi}-1\right\}\ 
.
\label{anR}
\end{equation}
When $R$ becomes large,
the coefficient $b_{n}(R)$ in the solution of scattering region behaves as
$
b_{n}(R)
\sim
e^{i(n-|\nu|/2)\pi}
\left\{{(kR)^{2}/ (4\alpha)}\right\}^{|\nu|/2}
/\Gamma(|\nu|+1)
$.
Recalling the 
definition of the Whittaker function and a relation between the 
hypergeometric functions
\begin{equation}
	\lim\limits_{\beta\rightarrow\infty}
	{}_{1}F_{1}(\beta;\gamma;z/\beta)
	={}_{0}F_{1}(\gamma;z),
\end{equation}
we observe
\begin{equation}
	{Rb_{n}(R)\over\sqrt{\alpha}r}
	M_{\lambda+\nu/2,\ |\nu|/2}(\alpha r^{2}/R^{2})
	\mathop{\longrightarrow}\limits_{R\rightarrow\infty}
	{(kr/2)^{|\nu|}\over\Gamma(1+|\nu|)}e^{i(n-|\nu|/2)\pi}
	{}_{0}F_{1}(1+|\nu|;-(kr/2)^{2})\ .
	\label{psi1n}
\end{equation}
By recognizing
\[
{(x/2)^{|\nu|}\over\Gamma(1+|\nu|)}{}_{0}F_{1}(1+|\nu|;-x^{2}/4)
=J_{|\nu|}(x)\ ,
\] 
we finally obtain the solution for the scattering region
\begin{equation}
 	\psi_{\text{I}}(r,\vphi)
 	=
 	\sum_{n=-\infty}^{+\infty}
	J_{|\nu|}(kr)e^{in\theta+i(n-|\nu|/2)\pi},
 	\label{toatalpsi}
\end{equation} 
as well as that for the asymptotic region
\begin{equation}
\psi_{\text{II}}(r,\vphi)
=
e^{ikr\cos\theta}+{1\over2}
\sum_{n=-\infty}^{+\infty}
e^{in(\theta+\pi/2)}\left\{e^{-i(|\nu|-n)\pi}-1\right\}
H_{n}^{(1)}(kr).
\label{psiasymp}
\end{equation}
 
Thus we have found that in the large $R$ limit  
the solution $\psi_{\text{I}}(r,\vphi)$ in the scattering 
region has the same form as the one that was obtained in 
the previous section through LS equation.
It should be noticed, however, that   
in this approach the wave function $\psi_{\text{II}}(r,\vphi)$ 
describes the scattering state in the 
asymptotic region far from the solenoid. 
From \eqref{psiasymp} we find that the incident 
wave is given by the plane wave and that 
the scattered wave is given by the second term of 
\eqref{psiasymp} denoted by $\psi_{\text{II,S}}(r,\vphi)$
\begin{equation}
	\psi_{\text{II,S}}(r,\vphi)=
	{1\over2}
	\sum_{n=-\infty}^{+\infty}
	e^{in(\theta+\pi/2)}\left\{e^{-i(|\nu|-n)\pi}-1\right\}
	H_{n}^{(1)}(kr).
	\label{psiscattgord}
\end{equation}
If we take the limit $r \to \infty$ of 
$\psi_{\text{II,S}}(r,\vphi)$, we 
again obtain the same scattering amplitude as in the previous 
section. Thus we conclude that 
the two approach to discuss the scattering problem of this system 
give the same result. 

Here it is better to give a comment on the relation between the 
methods explained here and in the previous section.
In finding the scattering amplitude in section \ref{LSanalysis}, 
information on the scattering has been given by the 
asymptotic behavior of the wave function that corresponds to 
$\psi_{\text{I}}(r,\vphi)$ in this section.
On the other hand, in this section, $\psi_{\text{II}}(r,\vphi)$ 
describes asymptotic behavior of the scattering state 
as has been shown above. The fact that these two methods give the 
same result is the consequence of the existence of the limit 
$R\rightarrow\infty$ in both $\psi_{\text{I}}(r,\vphi)$ and 
$\psi_{\text{II}}(r,\vphi)$ simultaneously. In other words, we may 
say that the Aharonov-Bohm scattering problem accepts the plane wave 
as a piece of its asymptotic wave function.
In this regard, we are reminded of the need of a more careful treatment 
in the same analysis of the scattering by the Coulomb potential.

\section{The scattering by a solenoid with a finite radius}
\label{gordon2}
As is easily seen from \eqref{wholepsi} or \eqref{toatalpsi}, the whole 
wave function of the scattering state neither vanishes nor is defined 
at the origin. Therefore the analyses in the preceding sections 
are unsatisfactory on this point. 
Fortunately the idea developed in the previous section is 
easily generalized to the system with a solenoid of finite radius. 
Repeating the same procedure with finite $r_{0}$, we obtain
\begin{eqnarray}
\psi_{\text{I}}(r,\vphi)
& = &
\sum_{n=-\infty}^{+\infty}
e^{in\theta+i(n-|\nu|/2)\pi}\left\{J_{|\nu|}(x)
-{J_{|\nu|}(a)\over H_{|\nu|}^{(1)}(a)}H_{|\nu|}^{(1)}(x)\right\},
\label{psiscgordf}\\
\psi_{\text{II}}(r,\vphi)
& = &
e^{ix\cos\theta}-{1\over2}
\sum_{n=-\infty}^{+\infty}
e^{in(\theta+\pi/2)}\left\{1+e^{-i(|\nu|-n)\pi}
{H_{|\nu|}^{(2)}(a)\over H_{|\nu|}^{(1)}(a)}\right\}
H_{n}^{(1)}(x),
\label{psiasympf}
\end{eqnarray}
where $a=kr_{0}$ and $x=kr$ and the wave function 
is assumed to vanish in the region $0<r \le r_{0}$. 
Again from the solution in the asymptotic region we 
can easily find the scattered wave 
\begin{equation}
	\psi_{\text{S}}(r,\vphi)=
	-{1\over2}
	\sum_{n=-\infty}^{+\infty}
	e^{in(\theta+\pi/2)}\left\{1+e^{-i(|\nu|-n)\pi}
	{H_{|\nu|}^{(2)}(a)\over H_{|\nu|}^{(1)}(a)}\right\}
	H_{n}^{(1)}(x),
	\label{psisf}
\end{equation}
and its asymptotic form for large $x$
\begin{eqnarray}
	\psi_{\text{S}}(r,\vphi)
	& \sim &
	{1\over\sqrt{r}}e^{ix-i\pi/4}f(\theta),
	\nonumber\\
	f(\theta)
	& = &
	-{1\over\sqrt{2\pi k}}
	\sum_{n=-\infty}^{+\infty}
	e^{in\theta}\left\{1+e^{-i(|\nu|-n)\pi}
	{H_{|\nu|}^{(2)}(a)\over H_{|\nu|}^{(1)}(a)}\right\}.
	\label{psisasympf}
\end{eqnarray}
The first term in the scattering amplitude exactly cancels the 
corresponding term from the incident plane wave. Therefore the 
$S$-matrix for the system is just a multiplication of a complex 
number 
of unit modulus:
\begin{equation}
	S_{n}=-e^{-i(|\nu|-n)\pi}
	{H_{|\nu|}^{(2)}(a)\over H_{|\nu|}^{(1)}(a)}\qquad
	(|S_{n}|=1),
	\label{smatn}
\end{equation}
on each eigenspace of the angular momentum. Hence the unitarity of 
the $S$-matrix is evident.  

Let us denote $\alpha=[\alpha]+\{\alpha\}$ again to write
\begin{equation}
	f(\theta)
	=
	-\sqrt{{2\over\pi k}}e^{-i[\alpha]\theta}\left\{
	e^{-i\pi\alpha/2}\sum_{n=0}^{\infty}
	e^{in\theta}A^{+}_{n}(\alpha)
	+e^{i\pi\alpha/2}\sum_{n=1}^{\infty}
	e^{-in\theta}A^{-}_{n}(\alpha)\right\},
	\label{psisasympf2}
\end{equation}
where $A^{\pm}_{n}(\alpha)$ is given in terms of the Bessel and the 
Neumann functions by
\begin{eqnarray}
	A^{+}_{n}(\alpha) & = & {1\over H_{n+\{\alpha\}}^{(1)}(a)}
	\left\{\cos(\pi\alpha/2)J_{n+\{\alpha\}}(a)
	-\sin(\pi\alpha/2)N_{n+\{\alpha\}}(a)\right\},
	\label{positc}  \\
	A^{-}_{n}(\alpha) & = & {1\over H_{n-\{\alpha\}}^{(1)}(a)}
	\left\{\cos(\pi\alpha/2)J_{n-\{\alpha\}}(a)
	+\sin(\pi\alpha/2)N_{n-\{\alpha\}}(a)\right\}.
	\label{negatc}
\end{eqnarray}
The total cross section is then found to be
\begin{equation}
	\sigma(\alpha)
	=
	{4\over k}\left\{
	\sum_{n=0}^{\infty}
	|A^{+}_{n}(\alpha)|^{2}
	+\sum_{n=1}^{\infty}
	|A^{-}_{n}(\alpha)|^{2}\right\}.
	\label{totalcs}
\end{equation}
This result explains an interesting feature of this system: 
$\sigma (\alpha)$ is apparently 
{\em periodic in $\alpha$ with period $2$\/}(not $1$). 

Unlike the case $r_{0}=0$(extremely thin solenoid), the wave function is 
strictly subjected to the boundary condition $\psi(r_{0},\vphi)=0$ 
even when $\alpha=\text{integer}$. 
Thus the solenoid is completely impenetrable to the charged particles.
Here let us consider the special case of $\alpha=\text{integer}$. 
The explicit form of $\sigma$ is found to be
\begin{equation}
	\sigma(\text{even})
	=
	{4\over k}{J_{0}^{2}(a)\over{J_{0}^{2}(a)+N_{0}^{2}(a)}}
	+{8\over k}\sum_{n=1}^{\infty}
	{J_{n}^{2}(a)\over{J_{n}^{2}(a)+N_{n}^{2}(a)}}
	\label{seven}
\end{equation}
for $\alpha=\text{even integer}$ and 
\begin{equation}
	\sigma(\text{odd})
	=
	{4\over k}{N_{0}^{2}(a)\over{J_{0}^{2}(a)+N_{0}^{2}(a)}}
	+{8\over k}\sum_{n=1}^{\infty}
	{N_{n}^{2}(a)\over{J_{n}^{2}(a)+N_{n}^{2}(a)}}
 	\label{sodd}
\end{equation}
for $\alpha=\text{odd integer}$.
When $a$ tends to $0$, these two formula behave in quite different 
ways. The formula \eqref{seven} is nothing but a total cross section 
of two dimensional hard core scattering, thus tends to $0$ with 
$a\rightarrow 0$ as $\pi^{2}/k\{\log(a/2)\}^{2}$. This result simply means 
that in the case of $\alpha=\text{even integer}$ charged particles are not 
affected by the solenoid at all. 
On the other hand, 
the formula \eqref{sodd} grows up to $\infty$ in the same limit since 
it has the Neumann function instead of the Bessel function in the 
numerator of each term. 
Therefore the total cross section of AB scattering for 
$\alpha=\text{odd integer}$ diverges 
when the radius of the solenoid tends to $0$.  
This singular behavior is the common feature of the total cross 
section except the case $\alpha=\text{even integer}$. If we notice 
that the partial cross section for large $n$ approaches immediately 
to $4\sin^{2}(\pi\alpha/2)/k$ even for finite $a$, we conclude that
the singularity is not the consequence of putting the radius of solenoid
infinitely small. 
This result implies the important fact that the vector potential can 
affect the charged particles even in the case of 
$\alpha=\text{odd integer}$, which is the 
different conclusion from that of AB.  
 
Apart from the divergence of the total cross section considered 
above, the unitarity of the $S$-matrix on each eigenspace of angular 
momentum is expected from \eqref{smatn}. 
This fact is also recognized from the different point of view.
According to the discussion given in the appendix \ref{appa}, 
the generalized optical theorem \eqref{unitaritya}, which is rewritten in 
terms of partial wave decomposition of a scattering amplitude
$f(\theta)=\sum_{n=-\infty}^{+\infty}e^{in\theta}f_{n}$ as
\begin{equation}
	|f_{n}|^{2}=-{1\over\sqrt{2\pi k}}(f_{n}+f_{n}^{*})\ ,
	\label{gopt1}
\end{equation}
is equivalent to the unitarity of the $S$-matrix.
In our problem $f_{n}$ is given by
\begin{equation}
	f_{n}=-\sqrt{2\over\pi k}A_{n\pm[\alpha]}^{\pm}(\alpha)
	e^{\mp i\pi\alpha/2}\ ,
	\label{fnab}
\end{equation}
where the upper and the lower signs correspond 
to $n+[\alpha]\ge 0$ or $n+[\alpha]< 0$, respectively.
Then the condition \eqref{gopt1} reads
\begin{equation}
	|A_{n}^{\pm}(\alpha)|^{2}
	=
	{1\over 2}\left\{e^{\mp i\pi\alpha/2}A_{n}^{\pm}(\alpha)
	+e^{\pm i\pi\alpha/2}A_{n}^{\pm\ *}(\alpha)\right\},
	\label{unitarypart}
\end{equation}
and is easily verified. 

As for the system with infinitely thin solenoid, an explicit form of 
the $S$-matrix is found and the operator identity 
$\hat{S}^{\dagger}\hat{S}=\hat{S}\hat{S}^{\dagger}=\bbox{1}$ can be 
verified directly. This is given in appendix \ref{smatrix}.

\section{Results and discussions}
\label{discuss}

In this article the scattering problem, 
first discussed by Aharonov and Bohm \cite{BoAh}, has been reconsidered. 
We have examined how the charged particle is scattered by the 
solenoid taking a plane wave as an incident wave in two ways ; 
by solving Lippmann-Schwinger equation ; 
by applying the Gordon's idea to the present situation. 
Furthermore the scattering problem by a solenoid of finite radius has 
been considered from the second viewpoint to ensure the 
impenetrability of the solenoid. 

We have shown that the two methods considered in this paper give the same 
wave function which represents the scattering state and that the 
results for the infinitely thin solenoid are exactly the same as was 
obtained by AB as far as the total wave function is concerned.  
However, it should be stressed again that 
the incident wave taken in this paper is
different from that of AB to result in the disagreement in the 
scattering amplitude of this paper to that of \cite{BoAh} or 
\cite{Takaba} by the term proportional to the $\delta$-function. 
It is the appearance of the $\delta$-function of 
forward direction that guarantees the unitarity 
of the $S$-matrix. On this point, our result agrees with that by 
Ruijsenaars \cite{Ruijs}. 
In appendix \ref{appb} we discuss the result of AB and give an 
explanation for the reason why we need the 
$\delta$-function in addition to the scattering amplitude given by AB.

We also have shown that the $\delta$-function term in the scattering 
amplitude makes the scattering cross section nonzero even in the case 
of $\alpha=\text{odd integer}$ as well as in the case of nonintegral $\alpha$.
It is better to make a comment for the case of integral $\alpha$. If we regard a change of variables
\[
(\psi(r,\vphi),\bbox{A}(\bbox{x}))
\mapsto
(\psi^{\prime}(r,\vphi),\bbox{A}^{\prime}(\bbox{x}))
=(e^{i\alpha\vphi}\psi(r,\vphi),\bbox{A}(\bbox{x})+{\hbar c\over e}{\alpha\over r}\bbox{e}_{\vphi})
\]
as a gauge transformation for integral $\alpha$, it seems that there is no AB effect for that case.
But it is not true
because this causes a change in the strength of the magnetic field at the origin. Furthermore the new wave function is not defined at the origin, which will brake the single-valuedness of the wave function. We may here refer the same argument for the vector potential given in \cite{PBAL}, \cite{PBALAS}.

\acknowledgments
The authors are grateful to thank Professor Y. Ohnuki and 
Professor S. Kamefuchi for many fruitful discussions.

\appendix
\section{Unitarity and optical theorem in two dimensional scattering}
\label{appa}
We provide in this appendix a note on two dimensional 
scattering theory for completeness.  
Suppose we have two solutions, 
$\psi_{k}(r,\vphi;\vphi_{0})$ and $\psi_{k}(r,\vphi;\vphi_{0}^{\prime})$, 
for 
a scattering problem corresponding to different incident beams of a same 
energy. They are assumed to have the asymptotic behavior
\begin{eqnarray}
	\psi_{k}(r,\vphi;\vphi_{0}) & \sim & 
	e^{ikr\cos\theta}+{1\over \sqrt{r}}e^{ikr-i\pi/4}f(\theta)\qquad
	(\theta=\vphi-\vphi_{0}),
	\label{asympu1}  \\
	\psi_{k}(r,\vphi;\vphi_{0}^{\prime}) & \sim & 
	e^{ikr\cos\theta^{\prime}}
	+{1\over \sqrt{r}}e^{ikr-i\pi/4}f(\theta^{\prime})\qquad
	(\theta^{\prime}=\vphi-\vphi_{0}^{\prime}),
	\label{asympu2}
\end{eqnarray}
where the phase factor $e^{-i\pi/4}$ has been introduced for later 
convenience.
If we assume the Hamiltonian to be Hermitian, it is straightforward 
to obtain
\begin{equation}
	\int_{-\pi}^{+\pi}d\vphi
	\left\{{\psi_{k}^{*}(r,\vphi;\vphi_{0}^{\prime})\derr
	\psi_{k}(r,\vphi;\vphi_{0})-\psi_{k}(r,\vphi;\vphi_{0})\derr
	\psi_{k}^{*}(r,\vphi;\vphi_{0}^{\prime})}\right\}=0
	\label{current}
\end{equation}
as a consequence of the Schr\"odinger equation. Taking $r$ 
sufficiently large and using the asymptotic form of the wave 
functions,
we immediately find
\begin{equation}
	\int_{-\pi}^{+\pi}d\vphi
	f^{*}(\vphi-\vphi_{0}^{\prime})f(\vphi-\vphi_{0})=
	-\sqrt{2\pi\over k}\left\{f(\vphi_{0}^{\prime}-\vphi_{0})
	+f^{*}(\vphi_{0}-\vphi_{0}^{\prime})\right\}.
	\label{unitaritya}
\end{equation}
This is the generalized optical theorem and is nothing but 
the $c$-number version of the unitarity of the $S$-matrix.
To see this, let us define $S$-matrix for the wave function given in 
\eqref{asympu1}.
From the asymptotic form of the 
wave function
 \begin{equation}
 	\psi_{k}(r,\vphi;\vphi_{0}) \sim 
	\sqrt{{2\pi\over kr}}\left[{
	ie^{-ikr+i\pi/4}\delta(\theta+\pi)
	+e^{ikr-i\pi/4}\left\{\delta(\theta)+\sqrt{k\over 2\pi}f(\theta)\right\}
	}\right],
 	\label{defsmat}
 \end{equation} 
we can find a definition of operators $\hat{S}$ and $\hat{f}$
\begin{equation}
	\hat{S}={\bf 1}+\hat{f},\quad
	(\hat{S}F)(\vphi)=F(\vphi)
	+\sqrt{k\over 2\pi}
	\int_{-\pi}^{+\pi}d\vphi_{0}f(\vphi-\vphi_{0})F(\vphi_{0}).
	\label{defsop}
\end{equation}
Then unitarity of the operator $\hat{S}$ reads
\begin{equation}
		\hat{f}^{\dagger}\hat{f}=
		-(\hat{f}+\hat{f}^{\dagger}),
		\label{unitarityop}
\end{equation}
which is equivalent to \eqref{unitaritya}.
 
As a special case of \eqref{unitaritya} or \eqref{unitarityop}, 
we can easily obtain the optical theorem just by putting 
$\vphi_{0}=\vphi_{0}^{\prime}$
\begin{equation}
	\sigma=-\sqrt{2\pi\over k}2\text{Re}(f(0)).
	\label{opti}
\end{equation}

\section{$S$-matrix of the AB scattering}
\label{smatrix}
Taking a plane wave
\begin{equation}
	\langle\bbox{x}|\Phi(k,\vphi_{0})\rangle
	={1\over2\pi}e^{ikr\cos\theta}\qquad
	(\theta=\vphi-\vphi_{0})\ ,
	\label{planewave}
\end{equation}
as an eigenstate of the Hamiltonian($\hat{H}_{0}$) of a free 
particle, we obtain 
\begin{eqnarray}
	 \langle\bbox{x}|\Psi^{(+)}(k,\vphi_{0})\rangle
	 & = & 
	 {1\over2\pi}\sum_{n=-\infty}^{+\infty}
	 J_{|\nu|}(kr)e^{in\theta+in\pi-i|\nu|\pi/2}\ ,
	\label{psip}  \\
	\langle\bbox{x}|\Psi^{(-)}(k,\vphi_{0})\rangle 
	& = & 
	{1\over2\pi}\sum_{n=-\infty}^{+\infty}
	 J_{|\nu|}(kr)e^{in\theta+i|\nu|\pi/2}\ ,
	\label{psim}
\end{eqnarray}
as solutions of LS equations
\begin{equation}
	|\Psi^{(\pm)}(k,\vphi_{0})\rangle
	=|\Phi(k,\vphi_{0})\rangle+
	(E-\hat{H}\pm i\eps)^{-1}V|\Phi(k,\vphi_{0})\rangle\ .
	\label{lseqap}
\end{equation}
Here we again abbreviate $n+\alpha$ by $\nu$. A matrix element of $S$ 
operator is given in terms of $|\Psi^{(+)}(k,\vphi_{0})\rangle$ 
and $|\Psi^{(-)}(k,\vphi_{0})\rangle$ by
\begin{equation}
	\langle\Phi(p,\vphi_{p})|\hat{S}|\Phi(q,\vphi_{q})\rangle
	=\langle\Psi^{(-)}(p,\vphi_{p})|\Psi^{(+)}(q,\vphi_{q})\rangle\ .
\label{smatelem}
\end{equation}
Making use of the explicit form of 
$|\Psi^{(\pm)}(k,\vphi_{0})\rangle$, we can easily obtain
\begin{eqnarray}
	\langle\Psi^{(-)}(p,\vphi_{p})|\Psi^{(+)}(q,\vphi_{q})\rangle
	& = &
	\sum_{n,n'=-\infty}^{+\infty}{1\over(2\pi)^{2}}\int_{0}^{\infty}rdr
	\int_{-\pi}^{+\pi}d\vphi J_{|\nu'|}(pr)J_{|\nu|}(qr)
	\nonumber\\
	&	&
	\times 
	e^{-in'(\vphi-\vphi_{p})+in(\vphi-\vphi_{q})
	-i|\nu'|\pi/2+in\pi-i|\nu|\pi/2}
	\nonumber\\
	& = &
	{1\over\sqrt{pq}}\delta(p-q){1\over2\pi}\sum_{n=-\infty}^{+\infty}
	e^{in(\vphi_{p}-\vphi_{q})+2i\delta_{n}}\ ,
	\label{smat2}
\end{eqnarray}
where $\delta_{n}=(n-|\nu|)\pi/2$.
It is then straightforward to see
\begin{eqnarray}	
	\langle\Phi(p,\vphi_{p})|\hat{S}^{\dagger}
	\hat{S}|\Phi(q,\vphi_{q})\rangle 
	& = & 
	\int_{0}^{\infty}kdk\int_{-\pi}^{+\pi}d\vphi
	{1\over\sqrt{pk}}\delta(p-k){1\over\sqrt{kq}}\delta(k-q)
	\nonumber\\
	 &  & 
	\times{1\over(2\pi)^{2}}\sum_{n,n'=-\infty}^{+\infty}
	e^{-in(\vphi-\vphi_{p})-2i\delta_{n}}
	e^{in'(\vphi-\vphi_{q})+2i\delta_{n'}}
	\nonumber \\
	 & = & 
	 {1\over\sqrt{pq}}\delta(p-q)\delta(\vphi_{p}-\vphi_{q})\ .
	\label{sunitary}
\end{eqnarray}
In the same way $\hat{S}\hat{S}^{\dagger}=\bbox{1}$ can be verified. 
Therefore the $S$-matrix of the AB scattering is unitary. 
By performing the sum in \eqref{smat2}, we can further rewrite
\begin{equation}
	\langle\Phi(p,\vphi_{p})|\hat{S}|\Phi(q,\vphi_{q})\rangle
	=
	{1\over\sqrt{pq}}\delta(p-q)\left\{\cos\pi\alpha\delta(\theta)
	+i{\sin\pi\alpha\over\pi}e^{i[\alpha]\theta}P{1\over{e^{i\theta}-1}}
	\right\}\ .
	\label{smat3}
\end{equation}
Recalling the expression \eqref{psasymp1} for the scattering 
amplitude 
$f(\theta)$ given in section \ref{LSanalysis},
we find a fundamental operator relation
\begin{equation}
	\hat{S}=\bbox{1}+\hat{f}\ .
	\label{snadf}
\end{equation}
Furthermore, if we introduce common eigenstates of $\hat{H}_{0}$ and 
of the angular momentum by
\begin{equation}
 	|\tilde{\Phi}(k,n)\rangle=
 	{1\over\sqrt{2\pi}}\int_{-\pi}^{+\pi}d\vphi
 	e^{in\vphi}|\Phi(k,\vphi)\rangle\qquad
 	(n=0,\pm 1,\pm2,\ldots)\ ,
 	\label{fourietr}
\end{equation}
the operator $\hat{S}$ is diagonalized as
\begin{equation}
	\hat{S}=\int_{0}^{\infty}kdk\sum_{n=-\infty}^{+\infty}
	|\tilde{\Phi}(k,n)\rangle\langle\tilde{\Phi}(k,n)|
	e^{2i\delta_{n}}
	\label{sdiagon}
\end{equation}
to convince us that the solution of LS equation assures the 
unitarity of $S$-matrix as well as its commutability with 
$\hat{H}_{0}$.

\section{Result of AB and the unitarity of the $S$-matrix}
\label{appb}
\begin{figure}[p]
\epsfxsize=.75\hsize
\begin{center}
\leavevmode
\epsfbox{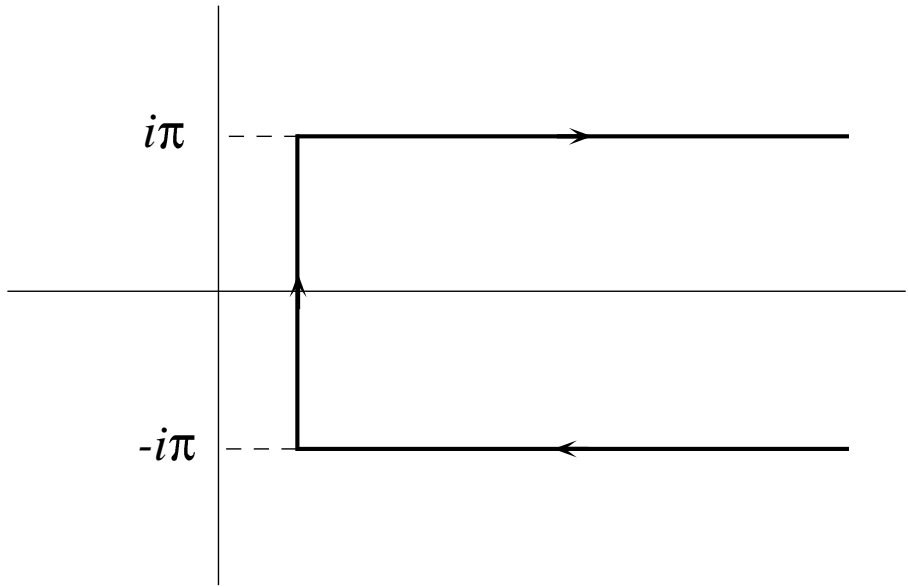}
\end{center}
\caption{Schl\"afli's contour of integral representaion of $J_{\nu}(x)$. 
$C:\ -i\pi+\infty\rightarrow +i\pi+\infty$.}
\label{fig1}
\end{figure}

\begin{figure}[p]
\epsfxsize=.75\hsize
\begin{center}
\leavevmode
\epsfbox{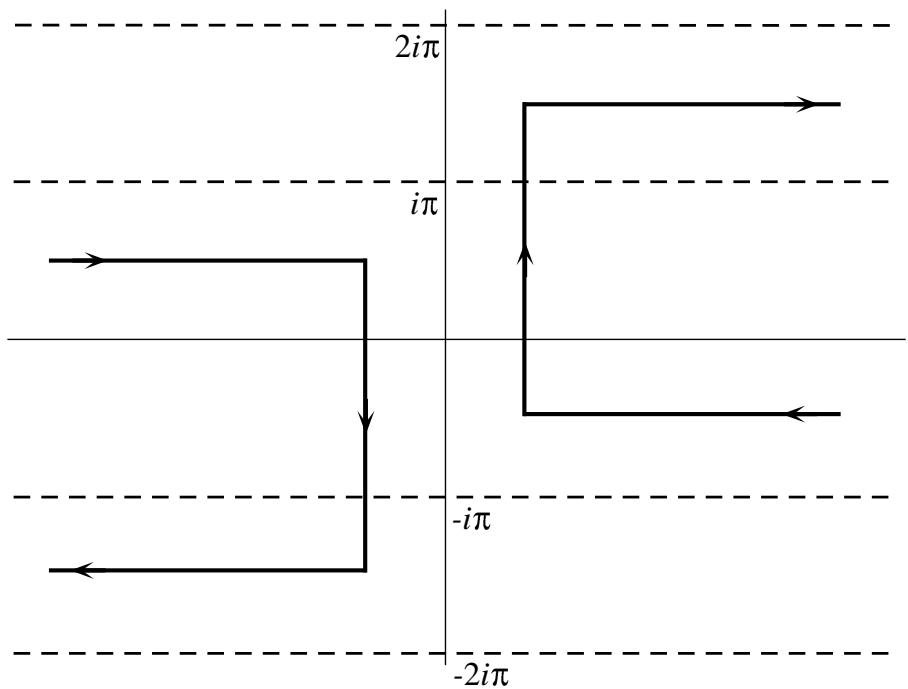}
\end{center}
\caption{Contours $C_{+}$ and $C_{-}$. 
$C_{+}:\ -i\pi/2+\infty\rightarrow +i3\pi/2+\infty$, 
$C_{-}:\ i\pi/2-\infty\rightarrow -i3\pi/2-\infty$}
\label{fig2}
\end{figure}

\begin{figure}[p]
\epsfxsize=.75\hsize
\begin{center}
\leavevmode
\epsfbox{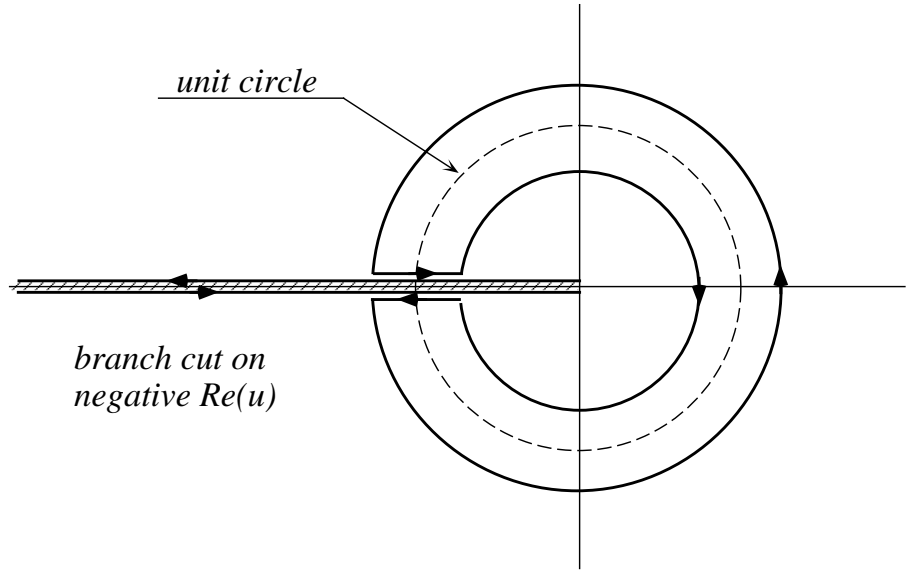}
\end{center}
\caption{Contour on $u$-plane corresponding to 
$|\Psi^{(+)}(k,\vphi_{0})\rangle$. There appears a branch cut on the negative $\text{Re}(u)$ axis. If $\theta=0$, we cannot adopt the residue theorem because the pole is located just on the branch cut.}
\label{fig3}
\end{figure}

\begin{figure}[p]
\epsfxsize=.75\hsize
\begin{center}
\leavevmode
\epsfbox{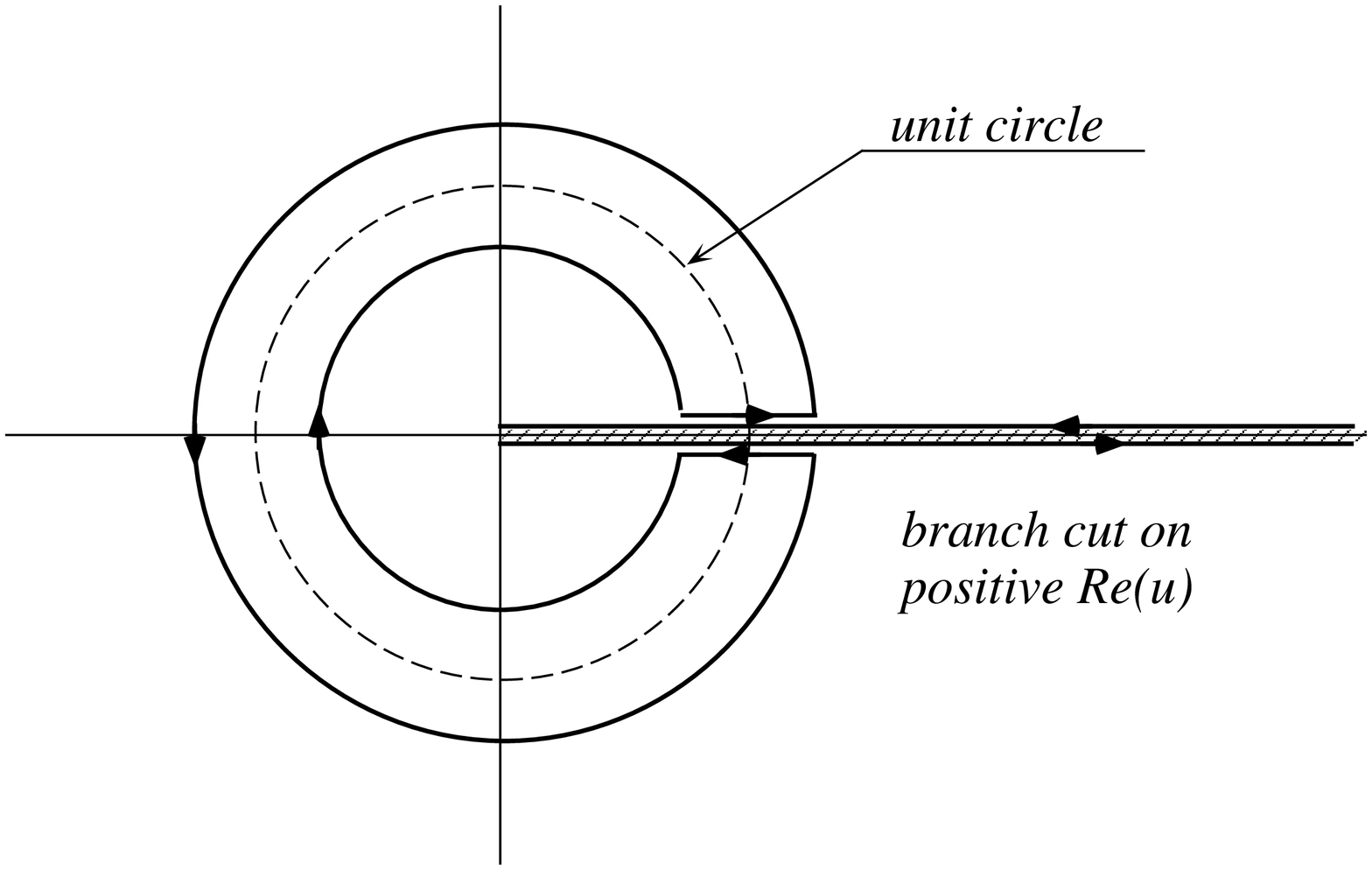}
\end{center}
\caption{Another contour on $u$-plane corresponding to 
$|\Psi^{(-)}(k,\vphi_{0})\rangle$. The branch cut appears in the opposite (backward) direction for this case.}
\label{fig4}
\end{figure}

The wave function of the scattering state is given by
\begin{equation}
	\psi_{\alpha}(r,\vphi)=\sum_{n=-\infty}^{+\infty}
	J_{|\nu|}(x)e^{in(\theta+\pi)-i|\nu|\pi/2},\qquad
	x=kr,\ \theta=\vphi-\vphi_{0},\ \nu=n+\alpha.
	\label{psitot}
\end{equation}
By use of the integral representation of the Bessel functions
\begin{equation}
	J_{\nu}(x)={1\over 2\pi i}\int_{C}dt
	e^{x\sinh t -\nu t}\qquad
	[\text{Re}(x)>0],
	\label{besselint}
\end{equation}
we can immediately convert \eqref{psitot} into its integral 
representation \cite{Stelit}, \cite{Jack}, \cite{Alva}
\begin{equation}
	\psi_{\alpha}={1\over 2\pi i}\int_{C}dt
	e^{x\sinh t}\left\{{e^{-\alpha t-i\pi\alpha/2}
	\over{1-e^{-t+i\theta+i\pi/2}}}
	+{e^{-(1-\alpha) t-i\theta+i(1+\alpha)\pi/2}
	\over{1-e^{-t-i\theta+i\pi/2}}}
	\right\},
	\label{psiint}
\end{equation}
for $0\le \alpha<1$. (See fig. \ref{fig1} for the contour $C$.) 
When $\alpha$ has an integral part 
($\alpha=[\alpha]+\{\alpha\}$), the wave function is obtained by 
$\psi_{\alpha}=e^{-i[\alpha](\theta+\pi)}\psi_{\{\alpha\}}$. We may, 
therefore, consider only the case of $0\le \alpha<1$. Making a change of 
variable, we can further rewrite \eqref{psiint} as
\begin{equation}
	\psi_{\alpha}=
	{1\over 2\pi i}\int_{C_{+}}dt
	e^{-ix\cosh t}{e^{(1-\alpha) t}
	\over{e^{t}+e^{i\theta}}}
	+{1\over 2\pi i}\int_{C_{-}}dt
	e^{-ix\cosh t}{e^{(1-\alpha) t}
	\over{e^{t}+e^{i\theta}}},
	\label{psiint1}
\end{equation}
where the contours $C_{+}$ and $C_{-}$ are depicted in fig. 
\ref{fig2}.
On change of variable $t\mapsto u=e^{t}$, there arises a multi-valued 
function $u^{-\alpha}$ in the integrand. Therefore we need to deal it 
with due care. If we recall that our solution for the scattering 
state, $|\Psi^{(+)}(k,\vphi_{0})\rangle$ in appendix \ref{smatrix}, 
has been obtained from the LS equation, 
we immediately notice that we have only one 
way to deform the contour to adopt the residue theorem to the 
integral on $u$-plane. (See fig. \ref{fig3}.) Another option for the 
deformation, fig. \ref{fig4}, obviously corresponds to another 
solution $|\Psi^{(-)}(k,\vphi_{0})\rangle$.  
As is seen from fig. \ref{fig3}, we can make use of 
the residue theorem to the contour integration 
around the unit circle only when $\theta\ne 0$. Then we obtain
\begin{equation}
	\psi_{\alpha}=
	e^{ix\cos\theta-i\alpha(\theta-{\scriptstyle{\rm sgn}}({\theta})\pi)}
	-{\sin\pi\alpha\over\pi}\int_{-\infty}^{+\infty}
	dt{e^{(1-\alpha)t}\over{e^{t}-e^{i\theta}}}
	e^{ix\cosh t}\qquad(\theta\ne 0),
	\label{psiint2}
\end{equation}
where
\begin{equation}
	\sgn{\theta}=
	\left\{{
	\begin{array}{cc}
	1&(0<\theta\le\pi)\\
	-1&(-\pi\le\theta<0)
	\end{array}
	}\right. .
\end{equation}
If we interpret the modulated plane wave
$\psi_{\text{inc}}
=e^{ix\cos\theta-i\alpha(\theta-{\scriptstyle{\rm 
sgn}}({\theta})\pi)}$
as an {\em incident wave}, the second term of \eqref{psiint2} will be 
regarded as a scattered wave. Then we will obtain the Aharonov-Bohm scattering 
amplitude $f_{\text{AB}}(\theta)$ 
with the aid of the stationary phase approximation from
\[
-{\sin\pi\alpha\over\pi}\int_{-\infty}^{+\infty}
	dt{e^{(1-\alpha)t}\over{e^{t}-e^{i\theta}}}
	e^{ix\cosh t}
	\sim{1\over\sqrt{r}}e^{ikr-i\pi/4}f_{\text{AB}}(\theta)\ .
\]
So far the amplitude $f_{\text{AB}}(\theta)$ has not been treated
in any connection with the $S$-matrix of the theory. Here it is important to note that we cannot define a $S$-matrix from \eqref{psiint2} because $\psi_\alpha$ in \eqref{psiint2} is not defined for $\theta=0$. Therefore it is inappropriate to decompose the total wave function in the form given above for considering the relation between the $S$-matrix and $f_{\text{AB}}(\theta)$. To find a definition of the $S$-matrix for this scattering problem, we need the asymptotic form of the total wave function
\begin{equation}
 	\psi_{\alpha}\sim
 	\sqrt{2\pi\over x}
	\left[e^{-ix+i\pi/4}\delta(\theta+\pi)
	+e^{ix-i\pi/4}\left\{
	\cos\pi\alpha\delta(\theta)
	+\sqrt{k\over2\pi}f_{\text{AB}}(\theta)
	\right\}\right]
 	\ .
 	\label{psitotasymp}
\end{equation} 
It should be then compared with \eqref{defsmat} and with discussion 
below in appendix \ref{appa}. For the present case, the $S$-matrix 
should be defined by
\begin{equation}
	\hat{S}=\cos\pi\alpha\bbox{1}+\hat{f}_{\text{AB}}\ ,
	\label{sabdef}
\end{equation}
which is nothing but the result given in \eqref{smat3}.
By equating both expressions in \eqref{sabdef} and that in 
\eqref{snadf}, we find 
\begin{equation}
	\hat{f}=(\cos\pi\alpha-1)\bbox{1}+\hat{f}_{\text{AB}}\ ,
	\label{fandfab}
\end{equation}
as the relation of the two scattering amplitudes.
Therefore $\int_{-\pi}^{+\pi}d\theta |f_{\text{AB}}(\theta)|^{2}$ 
cannot be interpreted as the total cross section. As a consequence, 
$f_{\text{AB}}(\theta)$  
does not obey the unitarity condition \eqref{unitarityop}. Rather, it 
satisfies an operator relation
\[
\hat{f}_{\text{AB}}^{\dagger}\hat{f}_{\text{AB}}=
\sin^{2}\pi\alpha\bbox{1}-
\cos\pi\alpha(\hat{f}_{\text{AB}}^{\dagger}+\hat{f}_{\text{AB}})
\]
because $S$-matrix itself has been shown to be unitary. In terms of the amplitude itself, it is expressed as
\[
\int_{-\pi}^{\pi}d\vphi f_{\text{AB}}^{*}(\vphi-\vphi_{f})
f_{\text{AB}}(\vphi-\vphi_{i})=
{2\pi\over k}\sin^{2}\pi\alpha\delta(\vphi_{f}-\vphi_{i})
\]
because $f_{\text{AB}}(\theta)$ satisfies
\[
f_{\text{AB}}^{*}(-\theta)+f_{\text{AB}}(\theta)=0\ .
\]

\begin{thebibliography}{3}
\bibitem{BoAh} Y. Aharonov and D. Bohm, Phys. Rev. {\bf 115}, 485 (1959).
\bibitem{Takaba} T. Takabayashi, {\em Hadronic Journal Supplement\/} 
{\bf 1}, 219  (1985).
\bibitem{PBAL} P. Bocchieri and A. Loinger, Nuovo Cimento, {\bf 47A}, 475(1978).
\bibitem{PBALAS} P. Bocchieri, A. Loinger and A. Siragusa, Nuovo Cimento, {\bf 51A}, 1(1979).
\bibitem{MVBer} M. V. Berry, Eur. J. Phys. {\bf 1}, 240 (1980).
\bibitem{OlaPop} S. Olariu and I. lovitzu Popescu, Rev. Mod. Phys. 
{\bf 57}, 339 (1985).
\bibitem{BerIno} C. C. Bernido and A. Inomata, J. Math. Phys. {\bf 22}, 
715 (1981). 
\bibitem{MorMen} G. Mornandi and E. Menossi, J. Phys. {\bf 5}, 49 
(1984)
\bibitem{GerSin} C. C. Gerry and V. A. Singh, Phys. Rev. D {\bf 20}, 
2550 (1979).
\bibitem{Shiekh} A. Y. Shiekh, Ann. Phys. {\bf 166}, 299 (1986).
\bibitem{Stelit} D. Stelitano, Phys. Rev. D {\bf 51}, 5876 (1995).
\bibitem{Ohnuki} Y. Ohnuki,  Proc. 2nd Int. Symp. Foundations of 
Quantum Mechanics, Tokyo, 1986,  pp. 117-126.
\bibitem{Gordon} W. Gordon, Zeits. f. Phys. {\bf 48}, 180 (1928).
\bibitem{Ruijs} S. N. M. Ruijsenaars, Ann. Phys. {\bf 146}, 1 (1983).
\bibitem{AALL} Y. Aharonov, C. K. Au, E. C. Lerner, and J. Q. Liang,
 Phys. Rev. D {\bf 29}, 2396 (1984).
\bibitem{Nagel} Bengt Nagel, Phys. Rev. D {\bf 32}, 3328 (1985).
\bibitem{Hagen} C. R. Hagen, Phys. Rev. D {\bf 41}, 2015 (1990).
\bibitem{Jack} R. Jackiw, Ann. Phys. {\bf 201}, 83 (1990).
\bibitem{Alva} M. Alvarez, hep-th/9510085.
\end{thebibliography}
\end{document}